\documentclass[aps,prb,twocolumn,superscriptaddress,nobibnotes,amsmath,amssymb,reprint]{revtex4-2}

\usepackage{graphicx}
\usepackage{xcolor}
\usepackage[normalem]{ulem}
\usepackage{natbib}

\begin{document}
\title{Comment on ``Shot noise in a strange metal''}
\author{B.A.~Polyak}
\affiliation{Osipyan Institute of Solid State Physics RAS, 142432 Chernogolovka, Russian Federation}
\author{E.S.~Tikhonov}
\affiliation{Osipyan Institute of Solid State Physics RAS, 142432 Chernogolovka, Russian Federation}
\author{V.S.~Khrapai}
\affiliation{Osipyan Institute of Solid State Physics RAS, 142432 Chernogolovka, Russian Federation}
\maketitle
The recent paper~\cite{doi:10.1126/science.abq6100} reports on the measurements of shot noise in the heavy fermion strange metal YbRh$_2$Si$_2$ patterned into the nanowire shape. The authors claim that the observed shot noise suppression can not be attributed to the electron-phonon energy relaxation in a standard Fermi liquid model but rather indicates the failure of quasiparticle concept. 

In this comment, we discuss the inconsistency in the resistivity extracted from nanowire devices of different length, which may indicate sizeable contribution of the contact resistance. The hint comes from the comparison of $660\,\text{nm}$-long device, presented in the main text, with the $30\,\mu\text{m}$-long device, presented in the Supplementary Materials (SM) and used for the determination of the electron-phonon coupling. Both nanowires are $60\,\text{nm}$-thick and their widths are close with the possible difference by few tens of percent. Still, the resistances are approximately $36\,\Omega$ for the shorter device and $255\,\Omega$ for the longer one (we cite the data at $3\,\text{K}$). The formally extracted resistivities of the two devices differ by a factor of approximately~$6$. Note that SM~Fig.S5 provides additional data on two more short nanowire devices with the same discrepancy. 

The authors of~\cite{doi:10.1126/science.abq6100} pay attention to the fact that nanoscale patterning may in principle change the underlying physics. They compare the temperature~($T$) and the magnetic field~($B$) behavior of the unpatterned film with that of the nanowire-patterned device and observe similar results in terms of the normalized resistance. However, in the nanowire-patterned device geometry the spreading resistance is in fact proportional to the YbRh$_2$Si$_2$ resistivity, 
\begin{equation*}
R_{\text{c}}\propto \rho\ln(\lambda/w),
\end{equation*}
where $w$ is the nanowire width and $\lambda$ is the current transfer length (between YbRh$_2$Si$_2$ and Au film, deposited on top) whose $T$-dependence is weak and closely follows $\lambda\propto\rho^{-1/2}$~\cite{tbp}. Therefore, similar $T$- and $B$-behavior of the unpatterned film with that of the nanowire-patterned device does not allow one to exclude the possible contribution of the contact resistance coming from the poor YbRh$_2$Si$_2$/Au interface. 

The possible contribution of spreading resistance is crucial in the analysis of shot noise suppression in the nominally short device. For the poor interface, $\lambda$ may by far exceed the length of the nanowire itself so that the significant voltage drops on the contact pads area, questioning the possibility to neglect the prosaic electron-phonon relaxation. We note that taking into account the spreading resistance allows us to consistently fit~\cite{tbp} the $T$-dependences of linear response resistance data for three short nanowire devices from Fig.2A and Fig.S5~\cite{doi:10.1126/science.abq6100} assuming the interface resistance of $200\,\text{k}\Omega/\mu \text{m}^2$ -- a high value which, however, can not be excluded based on the available data.

We thank Liyang Chen and Douglas Natelson for valuable comments on the device fabrication details. 
%
\end{document}